\documentclass[twocolumn,nofootinbib,amsmath,amssymb,aps,prd,balancelastpage,superscriptaddress]{revtex4-1}

\usepackage{amsmath,amssymb}
\usepackage{amsfonts,amssymb,mathrsfs}
\usepackage{graphicx}
\usepackage[bookmarks=false,pdfstartview=FitH]{hyperref}
\usepackage[all]{hypcap}
\usepackage{epsfig,latexsym}

\RequirePackage{xspace} \allowdisplaybreaks

\def\bef{\begin{figure}}
\def\eef{\end{figure}}

\newcommand{\be}[1]{\begin{equation}\label{#1}}
\newcommand{\beq}{\begin{equation}}
\newcommand{\ee}{\end{equation}}
\newcommand{\beqn}[1]{\begin{eqnarray}\label{#1}}
\newcommand{\eeqn}{\end{eqnarray}}
\newcommand{\bd}{\begin{displaymath}}
\newcommand{\ed}{\end{displaymath}}

\def\lsim{\raise0.3ex\hbox{$\;<$\kern-0.75em\raise-1.1ex
e\hbox{$\sim\;$}}}
\def\gsim{\raise0.3ex\hbox{$\;>$\kern-0.75em\raise-1.1ex
\hbox{$\sim\;$}}}
\def\simlt{\mathrel{\lower2.5pt\vbox{\lineskip=0pt\baselineskip=0pt
           \hbox{$<$}\hbox{$\sim$}}}}
\def\simgt{\mathrel{\lower2.5pt\vbox{\lineskip=0pt\baselineskip=0pt
           \hbox{$>$}\hbox{$\sim$}}}}
\def\unity{{\hbox{1\kern-.8mm l}}}
\def\lsim{\mathrel{\mathop  {\hbox{\lower0.5ex\hbox{$\sim$}
\kern-0.8em\lower-0.7ex\hbox{$<$}}}}}
\def\gsim{\mathrel{\mathop  {\hbox{\lower0.5ex\hbox{$\sim$}
\kern-0.8em\lower-0.7ex\hbox{$>$}}}}}

\def\be{\begin{equation}}
\def\ee{\end{equation}}

\usepackage{color}

\begin{document}

\pagestyle{plain}

\title{Gravitational waves from dark first order phase transitions and dark photons}

\author{Andrea Addazi}
\email{andrea.addazi@lngs.infn.it}
\affiliation{Department of Physics \& Center for Field Theory and Particle Physics, Fudan University, 200433 Shanghai, China}
\affiliation{Dipartimento di Fisica, Universit\`a di L'Aquila, 67010 Coppito AQ, Italy, EU\\
 LNGS, Laboratori Nazionali del Gran Sasso, 67010 Assergi AQ, Italy, EU}


\author{Antonino Marcian\`o}
\email{marciano@fudan.edu.cn}
\affiliation{Department of Physics \& Center for Field Theory and Particle Physics, Fudan University, 200433 Shanghai, China}

\begin{abstract}
\noindent
Cold Dark Matter particles may interact with ordinary particles through a dark photon, which acquires a mass thanks to a spontaneous symmetry breaking mechanism. We discuss a dark photon model in which the scalar singlet associated to the  spontaneous symmetry breaking has an effective potential that induces a first order phase transition in the early Universe. Such a scenario provides a rich phenomenology for electron-positron colliders and gravitational waves interferometers, and may be tested in several different channels. The hidden first order phase transition implies the emission of gravitational waves signals, which may constrain the dark photon's space of parameters. Compared limits from electron-positron colliders, astrophysics, cosmology and future gravitational waves interferometers such as eLISA, U-DECIGO and BBO are discussed. This highly motivates a {\it cross-checking strategy} of data arising from experiments dedicated to gravitational waves, meson factories, the International Linear Collider (ILC), the Circular Electron Positron Collider (CEPC) and other underground direct detection experiments of cold dark matter candidates.

\end{abstract}

\maketitle

\label{s.intro}

\noindent

\medskip

\section{Introduction}

\noindent
The possibility of testing first order phase transitions (FOPT) in the early Universe
seems to be more promising after the recent discovery of gravitational waves (GW) in LIGO
experiment \cite{Caprini:2015zlo,Kudoh:2005as}. In particular, next generations of interferometers like eLISA and U-DECIGO will be also fundamentally important 
to test gravitational signal produced by Coleman bubbles from FOPT. The production of GW from bubble collisions was first suggested in Refs.~\cite{Witten:1984rs,Turner:1990rc,Hogan:1986qda,Kosowsky:1991ua,Kamionkowski:1993fg}.

New experimental prospectives in GW experiments have motivated a {\it revival} of these ideas in context of new extensions of the Standard Model \cite{Schwaller:2015tja,Huang:2016odd,Artymowski:2016tme,Dev:2016feu,Katz:2016adq,Huang:2017laj,Baldes:2017rcu,Chao:2017vrq,Addazi:2016fbj}. In other words, the GW data may be used to test new models of particle physics beyond the standard model. 

In particular, contrary to electroweak FOPT, the presence of FOPT from a dark sector remains practically unconstrained. 

In this paper, we suggest to test/limit with GW experiments a minimal model of dark matter arising from a dark sector. Our proposal is based on the {\it dark photon} theory, 
first proposed by Holdom \cite{Holdom:1985ag}. In particular, we consider a hidden sector 
of a massive dark photon coupled to a massive dark fermion and a massive scalar field. 
The massive scalar spontaneously breaks the hidden electromagnetic symmetry, inducing a mass term for the dark photon. Now, the hidden scalar may undergo a violent first order phase transition for a large class of its effective self-interaction potentials. The spontaneously symmetry breaking process giving mass to the dark photon is highly motivated by  the strong constraints on long-range massless dark photons from orthopositronium experiments --- as first pointed out by Glashow \cite{G1,G2}. 

Our paper is organized as follows: in section II we will review some basics aspects of the massive dark photon model; in section III we will discuss the phenomenology of the model in GW interferometers and laboratory physics; in section IV we will spell out conclusions and remarks. 

\begin{figure}[t]
\centerline{ \includegraphics [height=6cm,width=0.8\columnwidth]{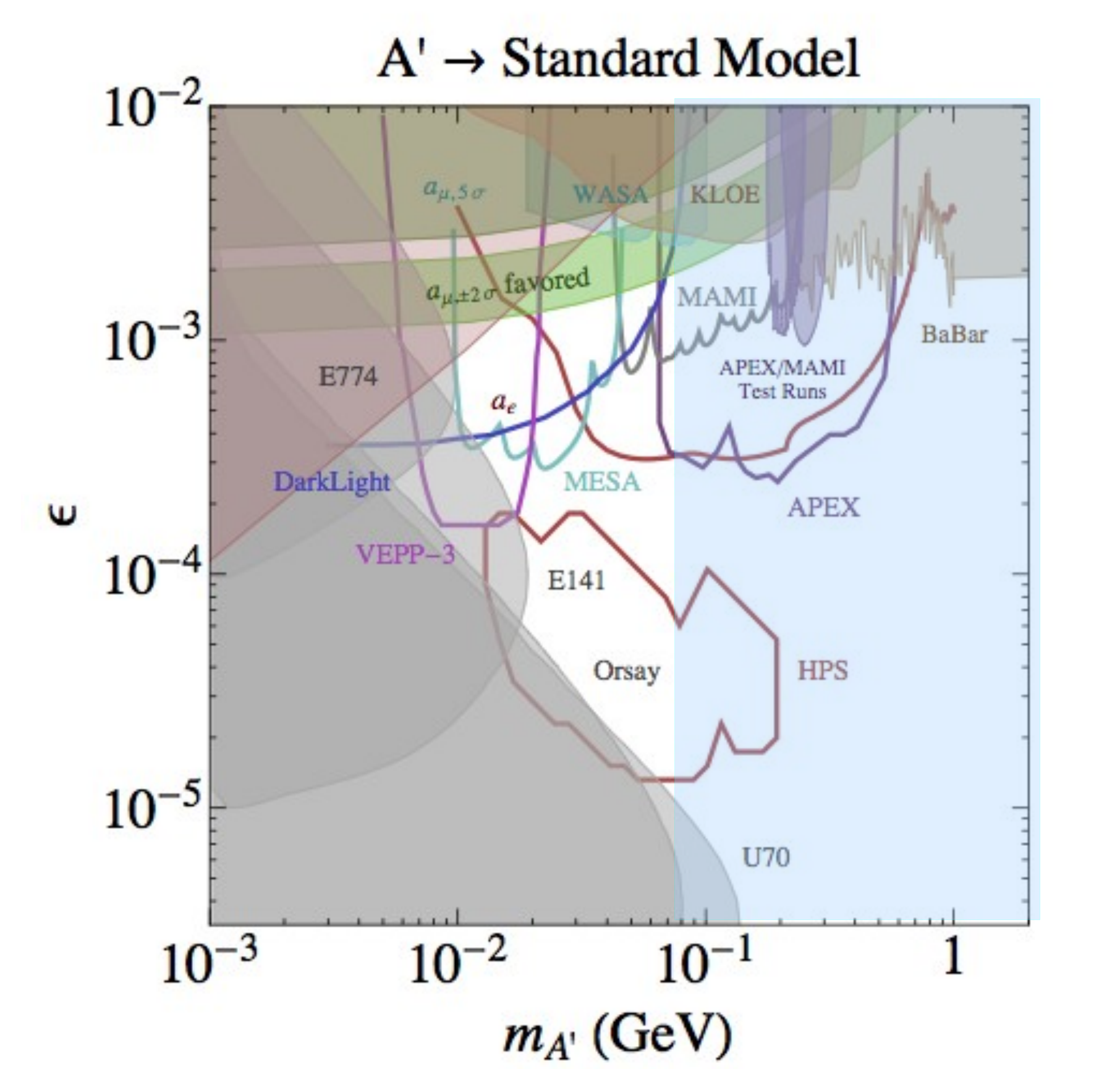}}
\vspace*{-1ex}
\caption{We show C.L. limits from SLAC and Fermilab experiments 
E137, E141, E774 \cite{116,117,119}, the electron and muon anomalous magnetic moment
$a_{\mu}$ \cite{120,121,122}, KLOE \cite{123,124}, WASA-at-COSY \cite{125},
APEX \cite{126} and MIAMI \cite{127}, BaBar \cite{116,128,129}
and supernova cooling constrains \cite{116,130,131}
--- 
APEX \cite{132}, HPS \cite{133}, DarkLight \cite{134}, 
VEPP-3 \cite{135,136}, MAMI and MESA \cite{137}
proposals are also reported. In light blue we show FOPT limits from future interferometers (eLISA, U-DECIGO and BBO \cite{Caprini:2015zlo,Kudoh:2005as,Audley:2017drz}) in the case of $\Lambda\, <\, 2.6$ TeV.}
\label{plot}   
\end{figure}

\begin{figure}[t]
\centerline{ \includegraphics [height=7cm,width=0.9\columnwidth]{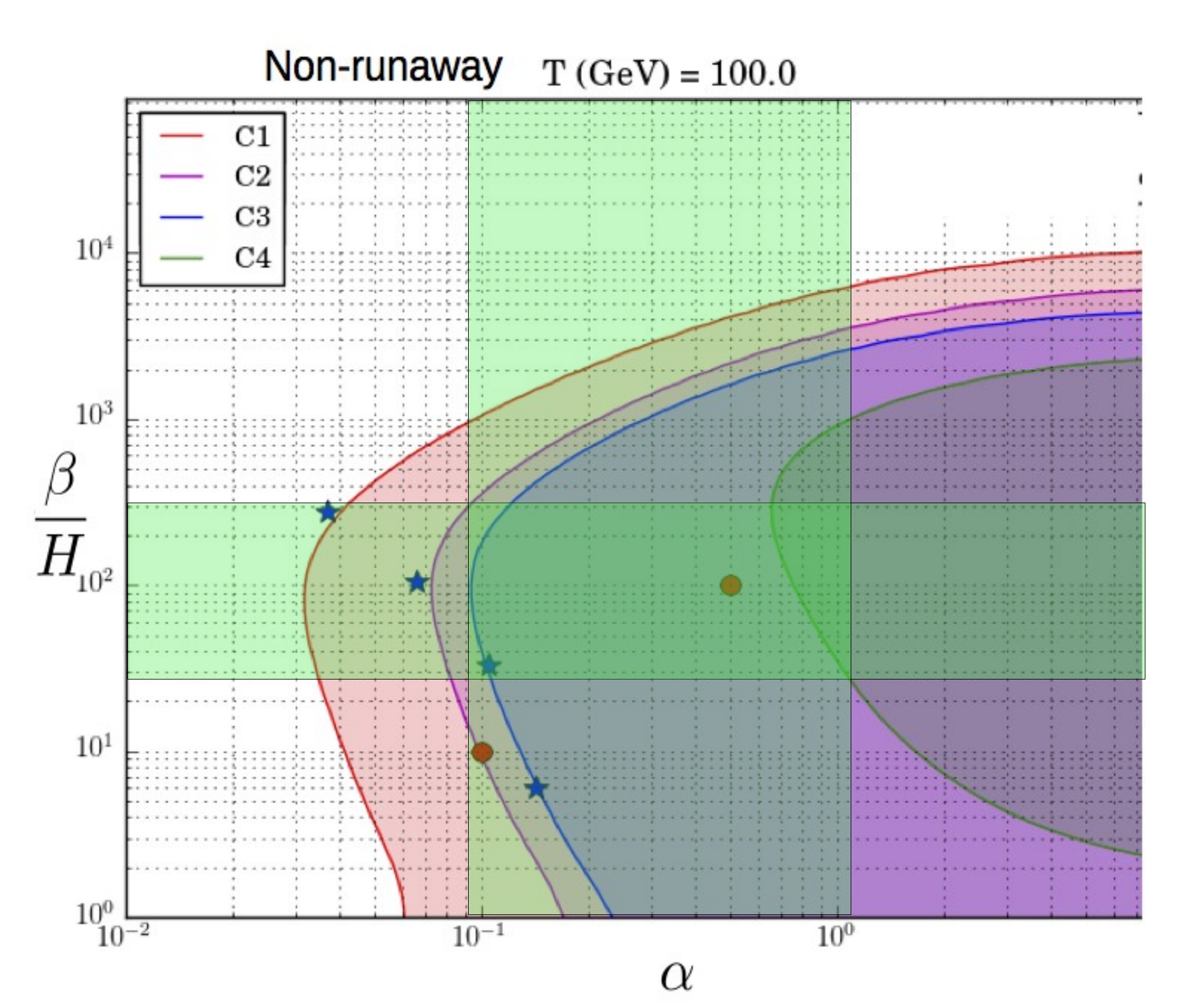}}
\vspace*{-1ex}
\caption{We show the predicted region for our model in the $(\alpha,\beta)$ parameters' space. This corresponds to the intersection of the two green regions, and is put in comparison with model independent regions for eLISA, as discussed in \cite{Audley:2017drz} assuming a VEV scale $100\, {\rm GeV}$. }
\label{plot}   
\end{figure}

\vspace{4cm}

\section{Dark photons model and first order phase transitions }
\noindent 
Let us consider the Standard Model extension with an extra abelian (non-anomalous) gauge $U(1)$, dubbed $U'(1)$, i.e. $SU_{c}(3)\times SU_{L}(2) \times U_{Y}(1)\times U'(1)$. The SM particles are assumed to be not charged with respect to such an extra $U'(1)$. Thus this latter singles out a dark abelian sector. Let us introduce a scalar singlet field $s$ and a Dirac fermion particle $\chi$, which are supposed to have a charge with respect to $U'(1)$, while the same are not charged --- thus are singlets --- with respect to the SM gauge group. In other words, the fermion and scalar are introduced as {\it hidden} particles. The dark gauge boson, dubbed dark photon $A'_{\mu}$, associated to $U'(1)$, may mix with the SM hypercharge boson $U(1)_{Y}$ through a renormalizable kinetic mixing term $-(\varepsilon/2) F^{Y}_{\mu\nu}F'^{\mu\nu}$ --- where $F_{\mu\nu}^{Y},F'_{\mu\nu}$ are respectively the field curvatures of the gauge bosons $Y_{\mu}$ and $A'_{\mu}$. The Lagrangian of the hidden sector reads
\noindent
\be \label{DL}
\mathcal{L}=K_{s}(s)+K_{\chi}(\chi)-\frac{1}{4}F'_{\mu\nu}F'^{\mu\nu}-\frac{\varepsilon}{2}F'^{\mu\nu}F^{Y}_{\mu\nu}+U(s,\chi)\,,
\ee
where
\noindent
\be \label{K}
K_{s}(s)+K_{\chi}(\chi)=(\mathcal{D}_{\mu}s^{\dagger})(\mathcal{D}^{\mu}s)+\bar{\chi}(i\gamma_{\mu}\mathcal{D}_{\mu}-\mu_{\chi})\chi
\ee
are hidden matter kinetic terms, $\mathcal{D}^{\mu}=\partial^{\mu}+ig'A'^{\mu}$ being the covariant derivative associated to the dark photon; $U(s,\chi)$ encodes the interactions of the hidden scalar and fermion fields:
\noindent
\be \label{UV}
U(s,\chi)=V(s)+y's\bar{\chi}\chi\,,
\ee
where $y'$ is a Yukawa-like free parameter and $V(s)$ is the singlet scalar self-interaction potential. In principle, the scalar singlet may interact with the SM Higgs field via the renormalizable interaction $\lambda_{sH}(s^{\dagger}s)(H^{\dagger}H)$. Such an interaction term may certainly provide an interesting portal to dark matter. However, the $\lambda_{sH}$ cannot be $O(1)$, otherwise the scalar singlet would not be hidden. Thus we will assume that the mixing term is highly suppressed. \\

The next main assumption of our model concerns the scalar singlet self-interaction potential. We assume that $V(s)$ drives the scalar field to get a VEV $\langle s \rangle=v_{s}$. In particular, we assume a double wells potential. On the other hand, we demand that the {\it wall} dividing the minima in the radial direction in the internal field space is lower than the standard quartic potential. In this way a highly unsuppressed first order phase transition is expected in the early Universe, as we will quantify in the following section. In particular, we will assume a simple effective potential of the form 
\noindent
\be \label{UV}
V(s)=m_{s}^{2}s^{\dagger}s+\frac{1}{4}\lambda_{S}(s^{\dagger}s)^{2}+\frac{1}{\Lambda^{2}}(s^{\dagger}s)^{3}+\cdots 
\ee

The main consequences of such a potential are the following:

i) The potential spontaneously breaks the $U'(1)$, giving a mass term to the dark photon $m_{A'}^{2}=g'^2 \, v_{s}^{2}$, where $v_{s}^{2}=-4m_{s}^{2}/\lambda_{s}$. 

ii) The potential will undergo a first order phase transition in temperature $\bar{T}\simeq v_{s}$. This may generate Coleman's bubbles, and bubbles-bubbles collisions generate a GW signal controlled by the scalar VEV-scale and the new physics scale $\Lambda$. 

iii) The dark matter particle is renormalized as $m_{\chi}=\mu_{\chi}+y'v_{s}$.  Eventually, we may assume that the bare mass is just zero and the dark matter mass is totally controlled by the singlet's VEV.

\section{Phenomenology}
\noindent
This simple minimal model leads to a rich phenomenology in several different channels. For instance, it allows multiple tests from particle physics experiments and gravitational waves interferometers. In the next section, we will start with a discussion of the GW signals that originate from the dark first order phase transition. Then, we will discuss how GW may test a region of parameters which may be confronted with limits from meson factories, electron-positron colliders and corrections to the magnetic moment of the electrons. 

\subsection{Gravitational waves signal}
\noindent
Let us remark that the frequency of the GW signal is controlled by the VEV scale of the first order phase transition. The frequency and the intensity of the gravitational waves signal 
have well known expressions, in which the model dependence enters only in the specification of the effective scalar field potential of the particular model considered\footnote{Recently, further numerical discussions
of GW productions from bubbles were shown in Refs.~\cite{Hindmarsh:2013xza,Hindmarsh:2015qta}.} \cite{Kamionkowski:1993fg}. 

The peak frequency of the GW signal produced by bubble collision has a value
$$\nu_{collision}\simeq 3.5 \times 10^{-4} \left(\frac{\beta}{H_{*}}\right)\left( \frac{\bar{T}}{10\, {\rm GeV}}\right)
\left(\frac{g_{*}(\bar{T})}{10}\right)^{1/6}\, {\rm mHz},$$
with corresponding intensity
$$\Omega_{collision}(\nu_{collision})\simeq $$
$$C\mathcal{E}^{2}\left(\frac{\bar{H}}{\beta}\right)^{2}\left(\frac{\alpha}{1+\alpha}\right)^{2}
\left( \frac{V_{B}^{3}}{0.24+V_{B}^{3}}\right)\left(\frac{10}{g_{*}(\bar{T})}\right).$$
In the latter relation we introduced 
$C\simeq 2.4\times 10^{-6}$, $$\mathcal{E}(\bar{T})=\left[T\frac{dV_{eff}}{dT}-V_{eff}(T)\right]_{T=\bar{T}},$$
\be \label{a}
\alpha=\frac{\mathcal{E}(\bar{T})}{\rho_{rad}(\bar{T})},\,\,\,\rho_{rad}=\frac{\pi^{2}}{30}g_{*}(T)T^{4}.
\ee
In eq.~\eqref{a} $\rho_{rad}$ stands for the radiation energy density, while $\bar{T}\simeq v_{s}$ denotes the first order phase transition temperature, defined by
\be \label{deba}
\beta=-\left[\frac{dS_{E}}{dt}\right]_{t=\bar{t}}\simeq \left[\frac{1}{\Gamma}\frac{d\Gamma}{dt}\right]_{t=\bar{t}},
\ee
in which 
$$S_{E}(T)\simeq \frac{S_{3}(T)}{T},\,\,\,\Gamma=\Gamma_{0}(T)\, {\rm exp}[-S_{E}(T)],$$
$$\Gamma_{0}(T)\sim T^{4},\,\,\,S_{3}\equiv \int d^{3}r\left(\partial_{i}s^{\dagger}\partial_{i}s+V_{eff}(s,T)\right).$$
The size of the bubble wall $\beta$, entering the definition in eq.~\eqref{deba}, is connected to the velocity of the bubble $V_{B}$ by the relation 
$$d\simeq \frac{V_{B}}{\beta}.$$

The tree-level effective potential is corrected by one-loop quantum corrections and thermal field theory corrections to $$V_{tree}(s,T=0)+V_{1}(s,T),$$ in which $$V_{1}(s,T)=V_{CW}(s,T=0)+\Delta V(s,T).$$
$V_{CW}$ is the one-loop Coleman-Weinberg potential, while $\Delta V(s,T)$ encodes thermal field theory contributions. The effective potential with a finite temperature --- similarly to the sixth order Higgs potential case studied in \cite{Delaunay:2007wb} --- 
can be approximated by 
$$V_{eff}(s,T)\simeq (m_{S}^{2}+CT^{2})s^\dagger s + \frac{\lambda}{4}(s^\dagger s)^{2} +\frac{1}{\Lambda^{2}}(s^\dagger s)^{3}\,,$$
where 
$$C=\frac{1}{4}\left(\pi \alpha'+\frac{m_{s}^{2}}{v_{s}^{2}}+y'^{2}-24\frac{v_{s}^{2}}{\Lambda^{2}}\right),$$
having introduced $\alpha'=g'^{2}/4\pi$. 

Further contributions are expected that arise from turbulence and sonic waves generated from the bubbles' expansion into the primordial plasma. Nonetheless these would only contribute for numerical prefactors in the estimate of the scale of the new physics involved, as shown in the following considerations. 

Assuming $\alpha'\!\sim\! \alpha\!\simeq\! 1/137$, the turbulence on the plasma induce by the bubble expansion may be estimated (see e.g. Refs.~\cite{Hindmarsh:2013xza,Hindmarsh:2015qta}) to have an expression $$\omega_{tur}\simeq O(1) \times 10^{-4}\left(\frac{\beta}{H_{*}}\right)\left(\frac{\bar{T}}{10\,{\rm GeV}}\right)\left(\frac{g_{*}}{10}\right)^{1/6}\,{\rm mHz},$$ $$\Omega_{tur}(\omega_{tur})\simeq O(1)\times 10^{-4}U_{T}^{5}V_{B}^{2}\left(\frac{H_{*}}{\beta}\right)^{2}\left(\frac{100}{g_{*}}\right)^{1/3},$$ where $U_{T}$ is the average ordinary and dark plasma velocity. 
We left a numerical prefactor undetermined, which traces back to an order $O(1)$ prefactor in the $\alpha'$.

We can now provide few estimates of orders of magnitude. In order to have a strong GW signal reachable by eLISA, U-DECIGO and BBO $$\Lambda/v_{s}\geq 24\div 26\,,$$
assuming $\alpha',y'\sim O(1)$. In particular GW frequencies scale with $T$, while the strain amplitudes scale as the inverse of $T$. Thus $v_{s}\sim 100\, \rm GeV$ with 
$\Lambda \sim 2.4\div 2.6\, {\rm TeV}$ corresponds to $\nu[Hz]\sim 10^{-1}\div 1\,{\rm mHz}$ (eLISA). The mass of the dark photon may be lowered with naturality 
by the gauge coupling $g'$ of $10^{-1}\div 10^{-3}$, in the interesting regime of dark photons
${\rm MeV}\div 10\, {\rm GeV}$.
Frequencies of $10^{-2}\div 10^{-3}\,{\rm mHz}$ correspond to scales of $v_{s}\sim 1\div 10\, {\rm GeV}$). A scale $v_{s}< 1\, {\rm GeV}$ is elusive to be detected in the minimal scenario. 

\subsection{Constraints on the dark photon}
\noindent 
Typically, the massive dark photon may have a mass $1\div 1000 \, {\rm MeV}$. Outside this range, the dark photon is very constrained by data. For instance, for a massless dark photon the kinetic mixing is $\sqrt{\alpha'}\varepsilon<10^{-7}$ from orthopositronium data with $m_{\chi}\simeq m_{e}$ \cite{G1,G2}. Also in a mass window $1\div 50\, {\rm MeV}$ the dark photon is very constrained. On the other hand, from $m_{A'}\sim 50\div 1000\, {\rm MeV}$, $\sqrt{\alpha'}\varepsilon$ may be high as $\sqrt{\alpha'}\varepsilon\sim 10^{-3}$. In Fig.1 compared constraints are displayed. 
Limits are mainly recovered from high luminosity low energy electron-positron colliders and astrophysics.

In the $1\div 1000 \, {\rm MeV}$ window of mass, the dark photon can be constrained by 
GW data in the framework of our model of a dark FOPT catalyzing the generation of the dark photon mass. Fixing various levels of the cutoff scale $\Lambda$, we can then superimpose the region in dark photon mass. The free relevant parameters for this model are $(\sqrt{\alpha'}\epsilon,m_{A'},v_{A'},\Lambda)$. The strategy may be then to fix $\Lambda$, and further impose constraints on the dark photon mass in the $(m_{A},\sqrt{\alpha'}\epsilon)$ parameters' space. As a result, GW tests result to be crucially important in order to get information on the dark Higgs sector generating the dark photon mass.

\section{Conclusions and remarks}
\noindent 
We discussed the possibility to test dark photon models from GW interferometers. 
In particular, the dark photon mass can be connected to a Higgs mechanism that undergoes to a FOPT in early Universe. We show that for dark photons of masses $10\div 1000\, {\rm MeV}$, eLISA, U-DECIGO and BBO interferometers may detect or rule-out dark FOPT related to it. 

We remark that our model leads also to an interesting phenomenology in Dark Matter direct detection experiments and new colliders. For example, a MeV-ish dark matter particle 
with a massive dark photon portal may interact mostly with electrons on DAMA detectors
 \footnote{See also  \cite{Addazi:2015cua,Cerulli:2017jzz} for other recent discussions of DAMA signal within the framework of dark gauge sectors.}  \cite{Bernabei:2007gr,Foot:2015vva,Lee:2015qva,Chen:2015pha}. So that the DAMA signal should be explained 
by energy recoils to electrons despite of nuclei, avoiding any detection by detectors like 
XENON and LUX --- these are not sensitive to those mass scales since electrons' signals  are cut in XENON/LUX double Xenon phase experiments. 

Another opportunity to detect over-GeV-ish dark photons might arise from the International Linear Collider (ILC) and Circular Electron Positron Collider (CEPC). In particular, they may be detected in missing transverse energy channels, which should be testable because of their high luminosity.


\noindent

\label{s.disc}

\noindent

\acknowledgments
\noindent
 
We 
thank USTC University for hospitality during the preparation of this paper. We 
are grateful to Arthur Kosowsky and Germano Nardini for enlightening discussions and remarks on these subjects. 

AM~wishes to acknowledge support by the Shanghai Municipality, through the grant No. KBH1512299, and by Fudan University, through the grant No. JJH1512105.

\end{document}